\documentclass[aps,prd,onecolumn,showpacs,reprint,nofootinbib,amsmath,amssymb,floatfix,superscriptaddress,showkeys]{revtex4}
\usepackage{graphicx}% Include figure files
\usepackage{dcolumn}% Align table columns on decimal point
\usepackage{bm}% bold math
\usepackage{captcont}
\usepackage{xcolor}
\usepackage{supertabular}

\usepackage{epstopdf}
\usepackage{mathtools}
\usepackage{natbib}
\usepackage{ulem}

\newcommand{\jcap}{J. Cos. Astropart. Phys.}

\begin{document}

\title{Search for a gamma-ray line feature from a group of nearby Galaxy clusters with Fermi LAT Pass 8 data}
\author{Yun-Feng Liang}
\affiliation{Key Laboratory of Dark Matter and Space Astronomy, Purple Mountain Observatory, Chinese Academy of Sciences, Nanjing 210008, China}
\affiliation{University of Chinese Academy of Sciences, Beijing, 100012, China}
\author{Zhao-Qiang Shen}
\affiliation{Key Laboratory of Dark Matter and Space Astronomy, Purple Mountain Observatory, Chinese Academy of Sciences, Nanjing 210008, China}
\affiliation{University of Chinese Academy of Sciences, Beijing, 100012, China}
\author{Xiang Li$^\ast$}
%\email{Corresponding author: xiangli@pmo.ac.cn}
\affiliation{Key Laboratory of Dark Matter and Space Astronomy, Purple Mountain Observatory, Chinese Academy of Sciences, Nanjing 210008, China}
\affiliation{University of Chinese Academy of Sciences, Beijing, 100012, China}
\author{Yi-Zhong Fan$^\ast$}
%\email{Corresponding author: yzfan@pmo.ac.cn}
\affiliation{Key Laboratory of Dark Matter and Space Astronomy, Purple Mountain Observatory, Chinese Academy of Sciences, Nanjing 210008, China}
\author{Xiaoyuan Huang$^\ast$}
%\email{Corresponding author: huangxiaoyuan@gmail.com}
\affiliation{Physik-Department T30d, Technische Universit\"at M\"unchen, James-Franck-Stra\ss{}e, D-85748 Garching, Germany}
\author{Shi-Jun Lei}
\affiliation{Key Laboratory of Dark Matter and Space Astronomy, Purple Mountain Observatory, Chinese Academy of Sciences, Nanjing 210008, China}
\author{Lei Feng}
\affiliation{Key Laboratory of Dark Matter and Space Astronomy, Purple Mountain Observatory, Chinese Academy of Sciences, Nanjing 210008, China}
%\affiliation{University of Chinese Academy of Sciences, Beijing, 100012, China}
\author{En-Wei Liang}
\affiliation{Guangxi Key Laboratory for the Relativistic Astrophysics, Department of Physics, Guangxi University, Nanning 530004, China}
\author{Jin Chang$^\ast$}
%\email{Corresponding author: chang@pmo.ac.cn}
\affiliation{Key Laboratory of Dark Matter and Space Astronomy, Purple Mountain Observatory, Chinese Academy of Sciences, Nanjing 210008, China}

\date{\today}

\begin{abstract}
Galaxy clusters are the largest gravitationally bound objects in the universe and may be suitable targets for indirect dark matter searches.
With 85 months of Fermi-LAT Pass 8 publicly available data, we analyze the gamma-ray emission in the directions of 16 nearby Galaxy Clusters with an unbinned likelihood analysis. No globally statistically-significant $\gamma-$ray line feature is identified and a tentative line signal may be present at $\sim 43$ GeV. The 95\% confidence level upper limits on the velocity-averaged cross section of dark matter particles annihilating into double $\gamma-$rays (i.e., $\langle \sigma v \rangle_{\chi\chi\rightarrow \gamma\gamma}$) are derived. Unless very optimistic boost factors of dark matter annihilation in these Galaxy Clusters have been assumed, such constraints are much weaker than the bounds set by the Galactic $\gamma-$ray data.
\end{abstract}


\maketitle

\section{Introduction}
In the standard $\Lambda$CDM cosmology model, the normal matter,
cold dark matter (DM), dark energy constitute about 5\%, 27\%, 68\% of the energy density of the today's universe, respectively.
DM is a new form of matter introduced to explain some gravitational effects observed in different scale structures such as the flat rotation curves of galaxies and the gravitational lensing of light by galaxy clusters that cannot be reasonably addressed by the amount of observed luminous matter \cite{Jungman1996,Bertone2005,Hooper2007,Feng2010}. Though much more abundant than the normal matter which can be exactly described within the standard particle physics model, the nature of DM is still unknown. Various hypothetical particles emerging in the extension of the standard particle physics model have been proposed to be the DM particles and the weakly interacting massive particles (WIMPs) are the leading candidates
\cite{Jungman1996,Bertone2005,Hooper2007,Feng2010}. Such kind of particles froze-out in the primordial
Universe, and this thermal production promises a non-negligible annihilation cross
section. If the interaction between these particles is in the electraweak scale, a velocity-averaged self-annihilation cross section of $\langle \sigma v \rangle \simeq 3 \times 10^{-26} \mathrm{cm}^{3} \;\mathrm{s}^{-1}$ would be expected, which would yield correct abundance of DM today.
Currently such a self-annihilation may still be efficient (for example, in the so-called s-wave annihilation scenario) and stable particles such as the electrons/positrons, protons/antiprotons, neutrinos/antineutrinos and gamma-rays are produced \cite{Jungman1996,Bertone2005,Hooper2007,Feng2010}. These particles are propagating into the space and could be sources of charged cosmic rays and diffuse gamma-rays. The main goal of DM indirect detection experiments is to distinguish between the DM annihilation (or decay) products and astrophysical background.

Historically, the kinematical study of the Coma cluster provided the first indication for the existence of DM \cite{Zwicky1933}. As the largest gravitationally bound objects in the universe, Galaxy clusters (GCls) are one of the most attractive regions of interest for the people working on DM indirect detection. Cosmic rays originated from annihilation/decay of DM particles in Galaxy clusters are confined there and unable to reach the Earth while the $\gamma-$rays can. Such $\gamma-$rays may have a line-like spectrum (double or even triple lines are also possible, depending on the annihilation final states and the rest mass of the DM particles) superposed by a continuous spectral component \cite{Bringmann2012Review} and their spatial distribution is expected to follow that of DM particles (if the number of signal photons is very limited, the statistical fluctuation effect should be taken into account \cite{Yang2012}). With 6 years of COMPTEL data collected during the extended observational programme of the Compton Gamma Ray Observatory, Iyudin et.al. \cite{Iyudin2004} carried out the first line search from a few very nearby GCls.

Since the successful launch of the {\it Fermi} Gamma Ray Space Telescope in June 2008 \cite{Atwood2009}, dedicated searches on possible DM annihilation/decay signal from GCls have been continually carried out. With the 11 months of Fermi-LAT data, Ackermann et al. \cite{Ackermann2010} searched for DM annihilation signals from GCls and the null results were taken to derive limits on the DM annihilation rates for the channels of $\chi\chi \rightarrow \bar{b}b$ and $\chi\chi \rightarrow \mu^{-}\mu^{+}$. The null results were also adopted to set limits on the DM decay rates \cite{Dugger2010}. The constraints however are uncertain since the annihilation signal can be significantly boosted due to the presence of DM substructures that however are still in debate \cite{Sanchez-Conde2011,Pinzke2011,Gao2012,Sanchez-Conde2014}.
Based on three years of Fermi-LAT gamma-ray data, Huang et al. \cite{Huang2012a} analyzed the flux coming from eight nearby clusters individually as well as in a combined likelihood analysis and imposed tight constraint on the annihilation and decay channels. In a joint likelihood analysis searching for spatially extended gamma-ray emission at the locations of 50 GCls in four years of Fermi-LAT data, no significant gamma-ray emission was obtained yet \cite{Ackermann2014GC}.

Among possible DM indirect detection signals, gamma-ray line(s), if not due to the instrumental effect, is believed to be a smoking-gun signature since no known physical process is expected to be able to produce such a kind of spectral feature(s). The branching fraction of mono-energetic DM annihilation channels is typically loop-suppressed and $\langle\sigma v\rangle_{\chi\chi\rightarrow \gamma\gamma} \sim (10^{-4}-10^{-1})\langle \sigma v \rangle$, where $\langle\sigma v\rangle_{\chi\chi\rightarrow \gamma\gamma}$
 is the cross section for DM particle annihilation into
a pair of $\gamma-$rays \cite{Bergstrom1997}. In 2012, possible evidence for the presence of a $\sim 130$~GeV $\gamma-$ray line signal in the inner Galaxy had been suggested \cite{Bringmann2012,Weniger2012,Tempel2012,Su2012,Yang2013}. Hektor et al. \cite{Hektor2012} reported further though a bit weaker evidence for the $\sim 130$~GeV $\gamma-$ray line emission from galaxy clusters in Fermi-LAT data (see however \cite{Huang2012b,Anderson2016}). The later analysis and in particular the latest analysis of the Pass 8 Fermi-LAT data for the Galactic center however does not confirm the presence of $\sim 130$~GeV $\gamma-$ray line feature \cite{Ackermann2013Line,Albert2015Line}. The search for line signal in the five years of the Fermi-LAT P7Rep data of 16 GCls also yielded null results \cite{Anderson2016}. Different from all previous related studies on GCls, in this work we analyze the publicly available Pass 8 Fermi-LAT data ranging from 27 Oct 2008 to 27 Nov 2015 and at energies between 1 and 300~GeV, in particular the sub-classes with improved energy resolution that is expected to enhance the line search sensitivity significantly. The main purpose of this work is to examine whether some unexpected spectral signals present in the latest Pass 8 data of some GCls that are selected from the extended HIghest X-ray FLUx Galaxy Cluster Sample (HIFLUGCS) catalog of X-ray flux-limited GCls  \cite{Reiprich2002,Chen2007}.

\section{Data Analyses}
\label{sec_data}
\subsection{Data selection}
The newly released Pass 8 data (P8R2 Version 6) from the Fermi Large Area Telescope (LAT)\cite{atwood09lat} is used in the present work.  The Pass 8 data provides a number of improvements over the previous Pass 7 data, including the better energy measurements, wider energy range and larger effective area \cite{pass8econf}. For the ``CLEAN" data, the effective area in Pass 8 increases by ${\sim}$30\% for events above 10 GeV \cite{fermi2015line}. The Pass 8 data have also been further divided into different \textit{event types} based on the energy  reconstruction quality with corresponding instrument response functions (denoted by EDISP0${\sim}$EDISP3  with larger number indicating better data quality, where EDISP represents ``energy dispersion"). The line search will be considerably benefitted from both the improved effective area and better energy resolution by using just the high quality data.

We take into account 85 months (from 2008-10-27 to 2015-11-27, i.e. MET 246823875 - MET 470288820\footnote{Data before MET 246823875 have a significantly higher level of background contamination at energies above $\sim$30 GeV and are not be used in our analysis (See \url{http://fermi.gsfc.nasa.gov/ssc/data/analysis/LAT_caveats.html}).}) of data, with the energies between 1 and 500 GeV.
We apply the zenith-angle cut $\theta < 90^\circ$ in order to avoid contamination from the earth's albedo, as well as the recommended quality-filter cuts (DATA\_QUAL==1 \&\& LAT\_CONFIG==1) to remove time intervals around bright GRB events and solar flares \footnote{\url{http://fermi.gsfc.nasa.gov/ssc/data/analysis/documentation/Cicerone/Cicerone_Data_Exploration/Data_preparation.html}.}. Except in Sec. III D, we make use of the ULTRACLEAN data set in order to reduce the contamination from charged cosmic rays. Since the energy resolution of EDISP0 data is much worse than that of the rest data and it just accounts for ${\sim}$1/4 of the whole data sets \footnote{\url{http://www.slac.stanford.edu/exp/glast/groups/canda/lat_Performance.htm}}, we exclude the EDISP0 data in most of our analysis to achieve better energy resolution and not significantly lose the statistics. The selection of events as well as the calculation of exposure maps are performed with the latest version of ScienceTools v10r0p5.

\subsection{Target clusters and binned stacking spectrum}
Our sample are the same as that of Anderson et al.\cite{Anderson2016}, which contains 16 GCls selected from the HIFLUGCS \cite{Reiprich2002, Chen2007}, including 3C 129, A 1060, A 1367, A 2877, A 3526, A 3627, AWM7, Coma, Fornax, M 49, NGC 4636, NGC 5813, Ophiuchus, Perseus, S 636 and Virgo. Among galaxy clusters whose parameters are reliably determined, these are the ones with largest $J$ factors.

Aperture photometry method is used to derive stacking spectral energy distribution (SED) of the sample consisting of 16 GCls.
Gamma-ray point sources are not expected to produce narrow line-like spectral features, so we do not mask any point sources around the target regions.
Angular radius of each ``region of interest" (ROI) is the radius corresponding to $R_{200}$ in Table.1 of \cite{Anderson2016}, where $R_{200}$ is the radius of a GCl inside which the average density is 200 times the critical density of the Universe $\rho_{\rm c}$ (Note that $\rho_{\rm c}=3 H_0^2 / 8 \pi {\rm G}$, and $H_0=67.79\ {\rm km\ s^{-1}\ Mpc^{-1}}$ \cite{planck2014}). Radii of ROIs of Virgo and M49 are taken as $2.6^\circ$ and $1.7^\circ$, respectively, to avoid the overlap between these two sources but keep the ratio between the ROI radii the same as that of $R_{200}$.
The stacking spectrum at energy $E_{\rm j}$ is derived by
\begin{equation}
(\frac{dN}{dE})_{\rm j} = \frac{\sum_{\rm i=1}^{16} n_{\rm ij}}{\bar{\epsilon}_{\rm j}{\Delta}E_{\rm j}\sum_{\rm i=1}^{16} \Omega_{\rm i}},
\end{equation}
where $n_{\rm ij}$ is the number of photons in each ROI at energy bin $E_{\rm j}$ , $\bar{\epsilon}_{\rm j}=\sum_{\rm i} {\Omega}_{\rm i}{\epsilon}_{\rm ij}/\sum_{\rm i} {\Omega}_{\rm i}$ is the averaged exposure weighted with solid angle $\Omega_{\rm i}$ at energy $E_{\rm j}$, and ${\Delta}E_{\rm j}$ is the width of given energy bin.
Fermi ScienceTools is used to select data within each ROI and calculate exposure maps.
Since the redshifts are all very small, we do not apply the redshift corrections to the spectrum.

The stacking SED based on aperture photometry is shown as red points in Fig.\ref{fig_sed_p8r2}.
At energies below $\sim 30$~GeV, the spectrum can be approximated by a power law, while at high energies, there is a cutoff. The high energy cutoff may be mainly due to the exponential cutoff in isotropic diffuse $\gamma$-ray background (IGRB) spectrum \cite{fermi2015iso}. Intriguingly, a possible spike structure appears at $\sim 43$~GeV, which is not expected in superposition of regular astronomical sources and motivates us to do the following further study. Please note that the binned stacking spectrum derived in this section is just for  ``visualization", and the following quantitative results do not rely on the binned analysis.

\subsection{Line fitting with unbinned likelihood method}

Since the binned stacking spectrum is sensitive to the adopted binning, we adopt an unbinned likelihood method to perform spectral fitting to further estimate the significance of the possible ``spike". The unbinned likelihood function is given by \cite{Ackermann2013Line}:
\begin{equation}
\ln\mathcal{L}(\lambda)=\sum_{\rm i=1}^N \ln(F(E_{\rm i};\lambda)\bar{\epsilon}(E_{\rm i})) - {\int}F(E;\lambda)\bar{\epsilon}(E)dE,
\end{equation}
where $N$ is the number of total $\gamma-$rays, $E_{\rm i}$ is energy of each $\gamma-$ray, and $F(E;\lambda)$ is the model flux with its variables $\lambda$, and $\bar{\epsilon}(E)$ is the exposure averaged over 16 GCls.

\begin{figure}[!h]
\includegraphics[width=0.7\columnwidth]{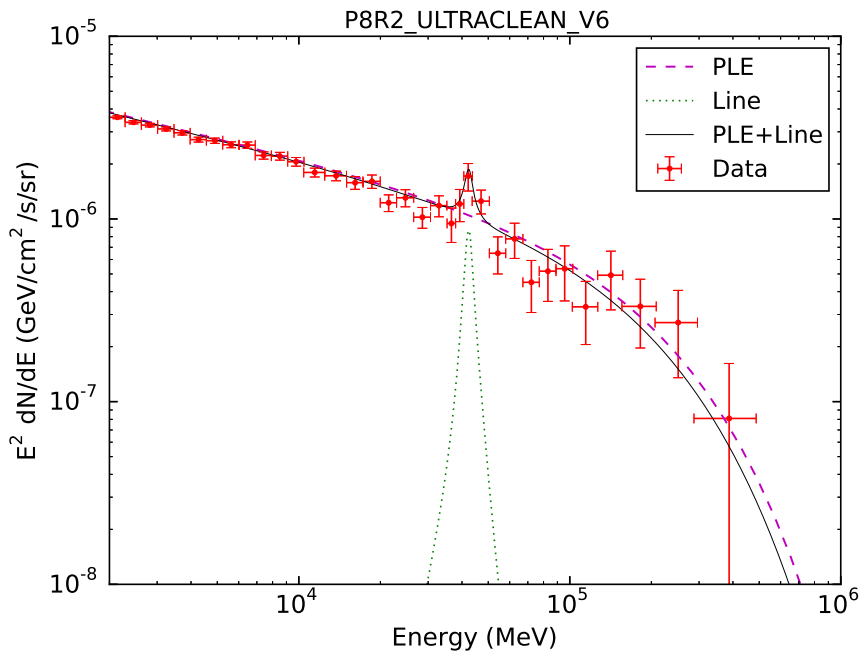}
\caption{The stacked spectral energy distribution of 16 galaxy clusters. Red points are the Fermi/LAT data and there might be a line-like structure at the energy of $\sim 43$ GeV (i.e., the dotted line).
}
\label{fig_sed_p8r2}
\end{figure}

Motivated by the presence of a high energy cutoff in Fig.\ref{fig_sed_p8r2}, we use a power-law with exponential cutoff (PLE) spectral function
\begin{equation}
F_{\rm b}(E;N_{\rm b},\Gamma,E_{\rm cut})=N_{\rm b}{\cdot}E^{-\Gamma}\exp{\left (-\frac{E}{E_{\rm cut}}\right )},
\end{equation}
to model the $\gamma-$ray background mixing point sources, galactic diffuse emission, isotropic component, and other components except a line signal.

We postulate that the signal is a monochromatic line (i.e. $S_{\rm line}(E)=N_{\rm s}{\cdot}\delta(E-E_{\rm line})$). Taking into account of the energy dispersion of Fermi LAT, the signal spectrum can be expressed as the following form,
\begin{equation}
F_{\rm s}(E;N_{\rm s},E_{\rm line})=N_{\rm s}{\cdot}D_{\rm eff}(E;E_{\rm line}),
\end{equation}
where $D_{\rm eff}$ is the exposure weighted energy dispersion function and is given by,
\begin{equation}
D_{\rm eff}(E;E')=\frac{\sum_{\rm k}\sum_{\rm j}{\epsilon}(E',{\theta}_{\rm j},s_{\rm k}) {\cdot} D(E;E',{\theta}_{\rm j},s_{\rm k})}{\sum_{\rm k}\sum_{\rm j} {\epsilon}(E',{\theta}_{\rm j},s_{\rm k})},
\end{equation}
$D$ is energy dispersion function of Fermi LAT\footnote{\url{http://fermi.gsfc.nasa.gov/ssc/data/analysis/documentation/Cicerone/Cicerone_LAT_IRFs/IRF_E_dispersion.html}}, and ${\epsilon}$ is exposure as function of incline angle respect the boresight $\theta$ and event type parameter $s$.

For a null hypothesis (non-signal hypothesis) and a signal hypothesis, the likelihood functions are
\begin{equation}
\ln\mathcal{L}_{\rm null}(N_{\rm b},\Gamma,E_{\rm cut})=\sum_{\rm i=1}^N \ln(F_{\rm b}(E_{\rm i})\bar{\epsilon}(E_{\rm i})) - {\int}F_{\rm b}(E)\bar{\epsilon}(E)dE,
\end{equation}
and
\begin{equation}
\ln\mathcal{L}_{\rm sig}(N_{\rm b},\Gamma,E_{\rm cut},N_{\rm s},E_{\rm line})=\sum_{\rm i=1}^N \ln(F_{\rm b}(E_{\rm i})\bar{\epsilon}(E_{\rm i})+F_{\rm s}(E_{\rm i})\bar{\epsilon}(E_{\rm line})) - {\int}(F_{\rm b}(E)\bar{\epsilon}(E)+F_{\rm s}(E)\bar{\epsilon}(E_{\rm line}))dE,
\end{equation}
respectively. Through maximizing likelihood of these two cases, we can obtain the best parameters describing the data, and calculate the test statistic (TS) value of the signal as
\begin{equation}
{\rm TS}\triangleq -2\ln\frac{\mathcal{L}_{\rm null}}{\mathcal{L}_{\rm sig}}.
\end{equation}

Our fit in energy range between 2 and 300~GeV (the range is little narrower than that of our entire data sets to allow for the spectral sidebands) displays a line at $E_{\rm line}=42.7{\pm}0.7~\mathrm{GeV}$ with a local test statistic value of $\rm{TS}=15.4$ (see Fig.~\ref{fig_sed_p8r2}).
MINUIT \cite{Minuit} is used in our fitting procedure. The black line in Fig.\ref{fig_sed_p8r2} represents the best fitting result. With the unbinned analysis result, we conclude that the excess is not an artificial product of binning.

%*****************************Fig.1***************************************
%*************************************************************************

\section{Testing possible origin of the excess signal}
\subsection{Sliding window analysis}
Furthermore, we use {\it sliding energy windows} technique \cite{Pullen2007, Abdo2010nc, Bringmann2012, Weniger2012,Anderson2016} to search for the $\gamma$-ray line like signal. This method avoids the bias caused by the inaccurate assumption of the background model (i.e., the PLE in above analysis). We choose a set of line energies $E_{\rm line}$ ranging from 5 GeV to 300~GeV. The interval of $E_{\rm line}$ between adjacent windows is roughly $0.5\sigma_{\rm e}$, where $\sigma_{\rm e}$ is the energy resolution (i.e., the 68\% energy dispersion containment half-windows)\footnote{\url{http://fermi.gsfc.nasa.gov/ssc/data/analysis/documentation/Cicerone/Cicerone_LAT_IRFs/IRF_E_dispersion.html}.}. Following \cite{fermi2015line}, the $E_{\rm line}$ of the first window is 5 GeV, and the size of each window is $E_{\rm line}\pm 0.5E_{\rm line}$. In such a narrow energy range, the gamma-ray background from diffuse and point sources could be approximated by a simple power law.
Unbinned likelihood method is used in each window (where $E_{\rm line}$ is fixed, and $f$ and $\Gamma$ are free), and thus we derive the TS value in each of these windows. This results is exhibited in Fig.\ref{slideWindowP8R2}. An excess emerges at $\sim 42$~GeV with the maximal TS value $\sim$16.7, corresponding to a local significance of about $4.1\sigma$.

\subsection{Earth limb}
The Earth limb is produced by interaction between cosmic-rays and the Earth's atmosphere. Such emission is peaked around zenith angle $Z{\sim}113^{\circ}$ and characterized by a soft spectrum of $dN/dE \propto E^{-2.8}$\cite{fermi09earth}.
The Earth limb has been widely adopted to examine the systematic effect of the instrument in previous studies \cite{fermi12line,Finkbeiner13lineSys,Ackermann2013Line} since the $\gamma$-rays resulting from atmospheric cascades are not expected to contain any line emission. For Fermi-LAT, the Earth limb is the brightest $\gamma$-ray source. Though with a soft spectrum, its counts rate is several times higher than any other astronomical sources even up to several hundreds of GeV. If the $\sim 42$~GeV line-like structure is due to an instrumental effect, for instance the anomalies of the energy reconstruction of gamma-ray events or bias of effective area in this energy range, it should cause distinct signal in the Earth limb data.

\begin{figure}[!h]
\includegraphics[width=0.43\columnwidth]{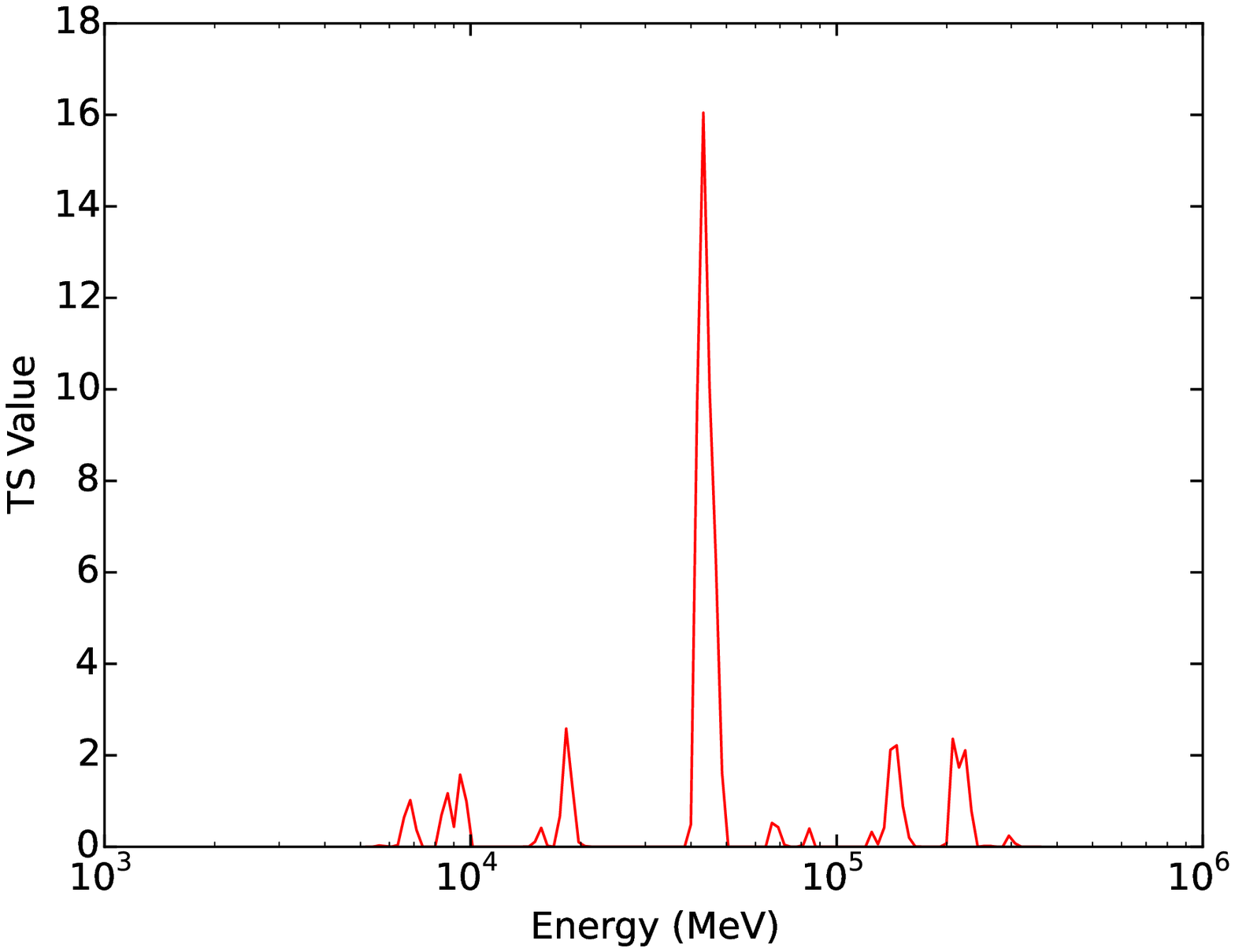}
\includegraphics[width=0.43\columnwidth]{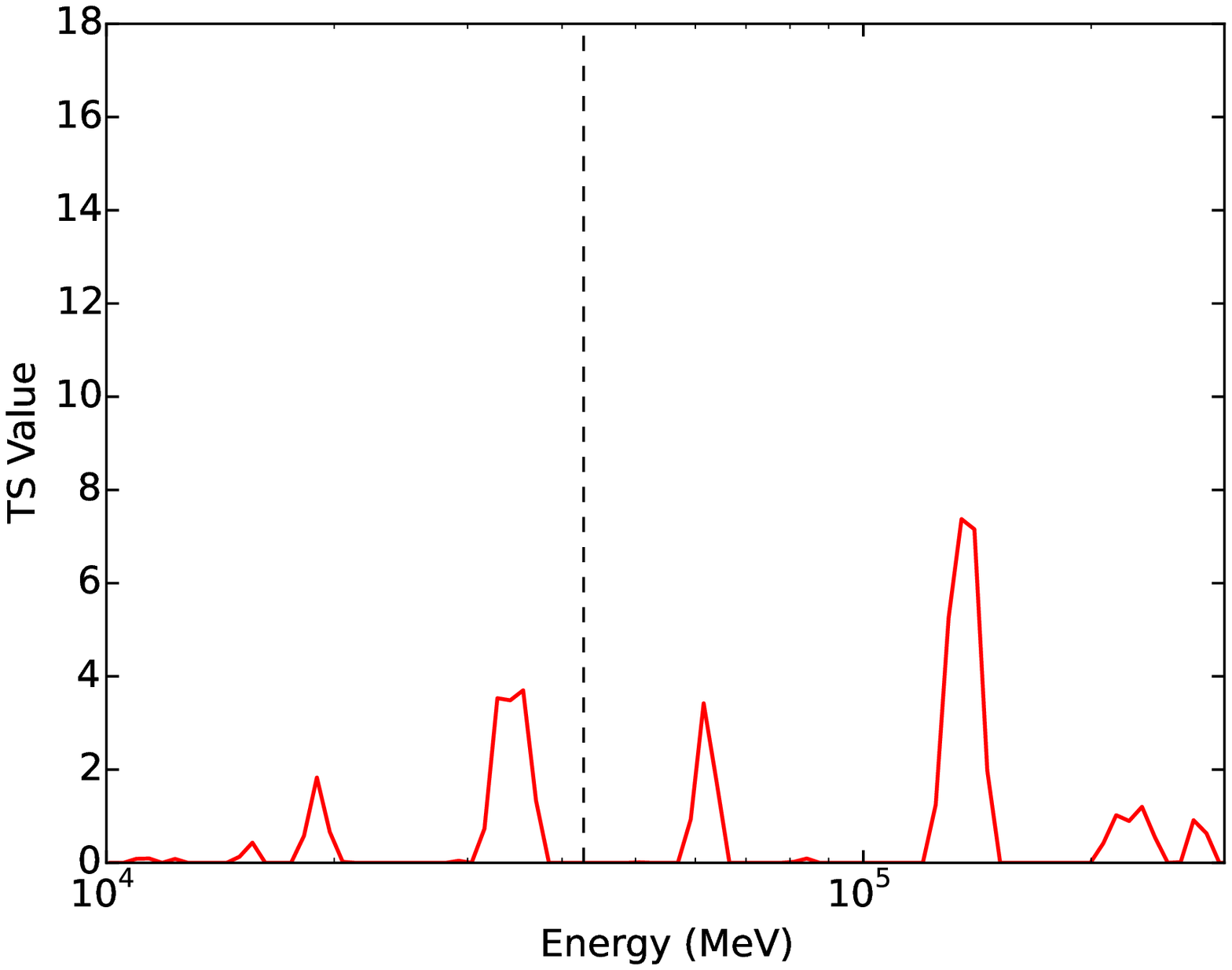}\\
\caption{Variability of TS value over a series of line energy in sliding window analysis of 16 GCls' (left panel) and the Earth limb's gamma-ray emission (right panel). In the left panel the peak with TS$\sim$16 appears at the energy $E_{\gamma}\approx 43$ GeV, while in the right panel no significant signal at such an energy is found.}
\label{slideWindowP8R2}
\end{figure}

Thus we selected photons within the zenith angles of $110^\circ-116^\circ$. We also restricted the rock angle of LAT instrument to be $>52^\circ$ to guarantee that the Earth limb photons have relative small incidence angles. Considering the fact that Earth limb is orders of magnitude higher than other astronomical sources \cite{fermi14earthlimb}, we simply use all the $\gamma-$rays passing these selection criterion.
Applying sliding window analysis on these earth limb photons, we did not find similar line-like signal in the data (see the right panel of Fig.\ref{slideWindowP8R2}).

%*****************************Fig.2***************************************
%*************************************************************************

%*****************************Fig.3***************************************
\begin{figure*}[t]
\begin{center}
\includegraphics[width=0.7\columnwidth,angle=0]{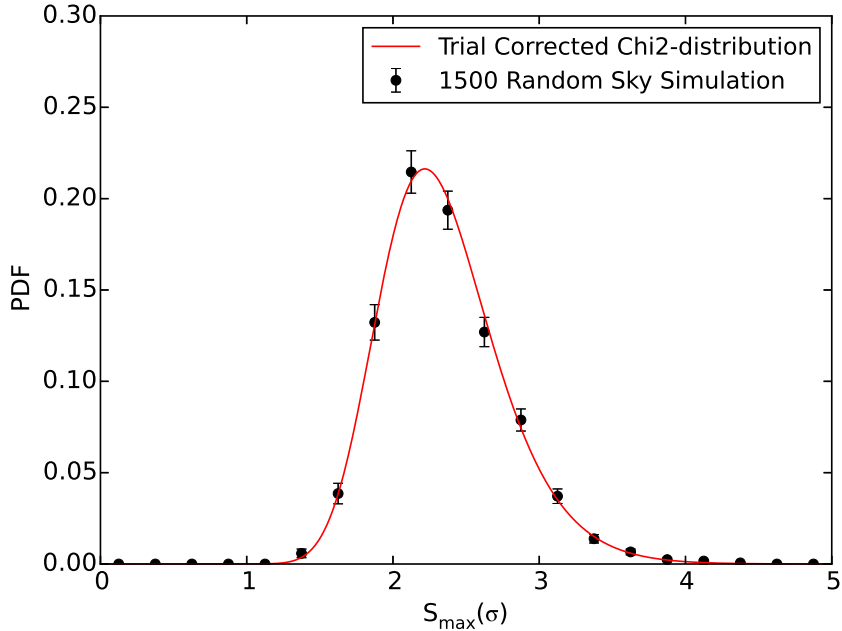}
\end{center}
\caption{Distribution of maximal TS value (${\sigma}_{\rm max}=\sqrt{{\rm TS}_{\rm max}}$) of 1500 random sky simulation (black points). The fit with a trial-corrected $\chi^2$ distribution (see the red curve) gives $k=0.97\pm0.19$ and $t=40.9\pm11.8$.}
\label{10sim}
\end{figure*}
%*************************************************************************

\subsection{Random sky simulations}
We also used random sky simulations to estimate the global significance of the possible signal. For each simulation a set of 16 ROIs are randomly selected. The above analysis on identifying line signal is reprocessed.
To emulate roughly the same environment of the GCl ROIs, we select 16 ROIs with the same radius as that of 16 GCls (i.e., for each Galaxy Cluster in our 16 samples there is a random-sky counterpart with the same radius).
The positions of simulated ROIs are randomly generated, but are constrained not lying within the regions of Galactic plane ($\left| b \right| < 15^\circ$) as well as the Galactic Center (i.e., $\left| b \right| < 20^\circ$ and $\left| l \right| < 20^\circ$), since majority of GCls are far away these two region\footnote{There are 3 GCls close to the Galactic plane, including 3C129, A3627 and Ophiuchus, the longitudes and latitudes of which are (160.43, 0.14), (325.33, -7.26)
and (0.56, 9.28), respectively. While removing these 3 sources from our sample and repeating the analysis in  Sec. II C we have
${\rm TS}=14.25$, implying that the potential signal is not a product of these 3 low latitude sources.} and Galactic Center is also a potential DM annihilation signal source.

In total 1500 sets of ROIs were generated, in each we carried out the sliding window analysis and recorded the resulting largest TS value. The distribution of these maximal TS values is shown in Fig.\ref{10sim}. We used a trial-corrected $\chi^2$ distribution \cite{trialfactor,Weniger2012}:
\begin{equation}
{\rm PDF}(\sigma_{\rm max};k,t) = \frac{d}{dx}{\rm CDF}(\chi^{2}_{\rm k};\sigma_{\rm max}^2)^t
\end{equation}
to fit the distribution and had $k=0.97\pm0.19$ and $t=40.9\pm11.8$, where $k$ is degree of freedom of $\chi^2$ distribution, $t$ is the number of independent trials, and ${\rm CDF}(\chi^{2}_{\rm k};\sigma^2)$ is the cumulative distribution function of $\chi^2$ distribution.
With this best fitted trial-corrected $\chi^2$ function, we have a global significance $\sim 3.0\sigma$ for TS $\approx 16.7$.

These simulations also disfavor the possibilities that the 42~GeV line-like structure is attributed to the analysis approach we used or alternatively it comes from a full sky isotropic component. This is because if the structure is caused by such possibilities, it will appear in the simulated spectra as well.

\subsection{Results based on other event classes/types}

For a monoenergetic line signal, it is expected to be more prominent in data set(s) with higher energy resolution. Benefited from improvements (namely \textit{event type}\footnote[7]{\url{http://fermi.gsfc.nasa.gov/ssc/data/analysis/documentation/Cicerone/Cicerone_Data/LAT_DP.html}}) in Pass 8 data, we can just take the photons with better energy reconstruction quality to test our results. Fermi-LAT data can also be separated into different \textit{event class} (SOURCE through ULTRACLEANVETO classes in Pass 8).
Data sets with higher probability of being $\gamma-$rays have lower contamination of background events, but smaller effective areas\footnotemark[7].

Now we carry out the same analysis procedure as Sec.\ref{sec_data} on data sets with different event classes/types. To have a reasonably-large statistics we take at least the sum of EDISP3 and EDISP2 data. The TS values together with number of events in energy range from 40GeV to 45GeV are summarized in Table \ref{tb1}. Indeed, a weak signal presents in all sets of data though the TS value changes (the maximal one has a TS$\sim 17$, similar to that found in Sec. III A).

%*****************************Tab.1***************************************
\begin{table}[ht]%[H] add [H] placement to break table across pages
\caption{TS value of the 43 GeV line-like signal.}
\begin{ruledtabular}
\begin{tabular}{lcccccccc}%c means center column...l/r for left/right
\multicolumn{1}{c}{} & \multicolumn{2}{c}{SOURCE} & \multicolumn{2}{c}{CLEAN} & \multicolumn{2}{c}{ULTRACLEAN} & \multicolumn{2}{c}{ULTRACLEANVETO}\\
\multicolumn{1}{c}{Event Type} & TS value &  Counts\tablenote{Number of events in the range from 40GeV to 45GeV}  & TS value & Counts  & TS value & Counts & TS value & Counts\\
\hline

FRONT+BACK\tablenote{i.e., EDISP(0+1+2+3)}              &  6.76   & 61 &  11.09    &  57 &   10.45  &  51  &  8.38    & 45   \\
EDISP(2+3)                                              &  10.61  & 38 &  13.80    &  34 &   13.06  &  30  &  15.31   & 28   \\
EDISP(1+2+3)                                            &  12.63  & 54 &  17.16    &  50 &   15.40  &  44  &  14.69   & 40   \\

\end{tabular}
\end{ruledtabular}
\label{tb1}
\end{table}
%*************************************************************************

\subsection{Constraints on $\langle \sigma v \rangle_{\chi\chi\rightarrow \gamma\gamma}$}

In the specific scenario of DM annihilation into a pair of $\gamma$-rays (i.e., $\chi\chi\to\gamma\gamma$), the flux from the combination of 16 GCls is given by
\begin{equation}
S_{\rm line}(E) = \frac{1}{4\pi} \frac{\left<\sigma v\right>_{\chi\chi\to\gamma\gamma}}{2m_\chi^2} \ 2\delta(E-E_{\rm line})\sum_{\rm i=1}^{16} J_{\rm i},
\label{eg:flux}
\end{equation}
where $m_\chi$ is the rest mass of the DM particle, $\left<\sigma v\right>_{\chi\chi\to\gamma\gamma}$ is the velocity-averaged annihilation cross-section for $\chi\chi\to\gamma\gamma$, $E_{\rm line}$ is the energy of mono-energetic photons which is $m_\chi$ here, and $J_{\rm i}$ is the $J$ factor of $i$-th GCl.

$J$ factor is concerned with the DM distribution along the line of sight within a given ROI, and is defined as
\begin{equation}
J = \int_{\rm ROI} d\Omega\int_{\rm l.o.s.}ds\ \rho(r(\rm s, \theta))^2,
\label{eg:j-factor}
\end{equation}
where $\rho$ represents the DM density distribution. In the current structure formation paradigm, GCls are formed through a hierarchical sequence of mergers and accretion of smaller systems \cite{Kravtsov2012}. Cosmological simulations show that a smooth host halo and a large number of sub-halos make up a cluster DM halo \cite{Gao2012}, and they are expected to be tightly related to strength of DM annihilation signal. Here we consider these two contributions separately.

We assume that the smooth halo follows a Navarra-Frenk-White (NFW) profile \cite{Navarro1997},
\begin{equation}
\rho_{\rm sm}(r) = \frac{\rho_0}{(r/r_{\rm s})(1+r/r_{\rm s})^2},
\end{equation}
where $r_{\rm s}$ denotes the scale radius and $\rho_0$ is the density normalization that are determined from the observational data \cite{Reiprich2002,Chen2007,Anderson2016}. We introduce the concentration parameter $c_{200} \equiv R_{200}/r_{\rm s}$. A relationship between the concentration parameter and the mass is shown by N-body simulation \cite{Gao2012}. Throughout this work we use the same $c_{200}$ as \cite{Anderson2016}, in which the concentration parameter of Virgo is taken from \cite{fermi2015Virgo} and others are calculated with the concentration-mass relation from \cite{Sanchez-Conde2014}. We obtain $r_{\rm s}$ using $R_{200}$ and $c_{200}$, and $\rho_0$ with $M_{200}$ (the mass of a GCl within the radius $R_{200}$). Then the $J$ factor of the smooth halo, $J_{\rm sm}$, is derived with eq.(\ref{eg:j-factor}).

The presence of DM sub-halos will make the annihilation rate enhanced (i.e., the boost factor $BF>1$) and the surface brightness profile less concentrated. However, current loose constraints on the sub-halo mass fraction, mass distribution function and concentration-mass relation make the value of $BF$ quite uncertain \cite[e.g.,][]{Gao2012,Sanchez-Conde2014,Gao2012b}.

The line signal shown in Sec. II C, if interpreted as the product of DM annihilation,
would suggest a $m_\chi \approx 42.7$ GeV and a $\left<\sigma v\right>_{\chi\chi\to\gamma\gamma}\approx 5\times 10^{-28}~{\rm cm^3~s^{-1}}(\overline{BF}/10^{3})^{-1}$, where $\overline{BF}$ is the poorly-constrained averaged boost factor of the DM annihilation of our GCl sample.

\begin{figure}[!h]
\includegraphics[width=0.7\columnwidth]{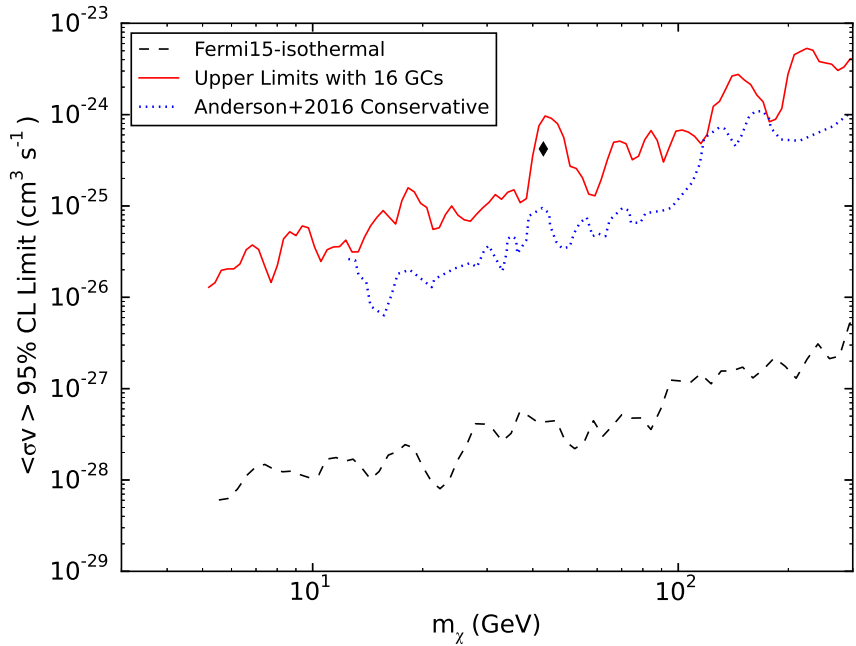}
\caption{The 95\% confidence level upper limits on the cross section of DM particles annihilating into double $\gamma$-rays obtained in our analysis of the 16 GCls (see the solid line) and that in the case of isothermal DM profile obtained by the Fermi-LAT collaboration \cite{fermi2015line} for the Galactic gamma-ray data (see the dashed line).
The filled diamond represents the required $\langle\sigma v\rangle_{\chi\chi\rightarrow \gamma\gamma}$ (no boost factor is introduced) for the possible $\gamma$-ray signal displayed in Fig. \ref{fig_sed_p8r2}.
The so-called ``conservative" result from \cite{Anderson2016} is also plotted for a comparison (see the dotted line). Note in the approach of \cite{Anderson2016}, boost factors around 30 for individual GCls have been adopted while in our approach no boost factor is assumed. Hence intrinsically (i.e., without introducing boost factors in both approaches) our constraints are (slightly) tighter than that of \cite{Anderson2016}. }
\label{fig_ul_anni}
\end{figure}

In view of the facts that the global significance is relatively low (i.e., $\sim 3.0\sigma$) and the instrumental effect is still to be fully  explored, as a conservative approach we calculate the upper limits on $\langle \sigma v \rangle_{\chi\chi\rightarrow \gamma\gamma}$ set by the $\gamma-$ray data of 16 GCls. We  increase $\left<\sigma v\right>_{\chi\chi\to\gamma\gamma}$ in eq.(\ref{eg:flux}) until the likelihood decreases by a factor of 1.35 with respect to the maximum one, and then obtain 95\% confidence level cross-section upper limit. In Fig.~\ref{fig_ul_anni} we present our constraints (without the introduction of boost factor) and that in the case of isothermal DM profile obtained by the Fermi-LAT collaboration \cite{fermi2015line} for a comparison. Evidently, the GCl constraints on $\langle \sigma v \rangle_{\chi\chi\rightarrow \gamma\gamma}$ is a few orders of magnitude weaker than the Galactic $\gamma-$ray data unless $BF \geq 10^{3}$ holds for the Galaxy Clusters in our sample. Such high $BF$s were proposed in \cite{Gao2012} but is still in debate  \cite{Sanchez-Conde2014}. If in reality $BF\ll 10^{3}$, the Galaxy Clusters are not compelling sources for DM indirect detection any longer.

%*****************************Fig.4***************************************
%*************************************************************************

\section{Discussion and Summary}
In this work, we have analyzed 85 month  publicly-available Pass 8 Fermi-LAT data (with energies between 1 and 300~GeV) in the directions of 16 Galaxy clusters selected from the extended HIFLUGCLS catalog of X-ray flux-limited sources that are expected to have large $J$ factors. Our main purpose is to search for any ``unusual" spectral signal displaying in the latest gamma-ray data. The weak gamma-ray line signal at energy of $\sim 130$ GeV from a group of Galaxy clusters, as reported in Hektor et al. \cite{Hektor2012}, is not found in our analysis (see the left panel of Fig.\ref{slideWindowP8R2} and also \cite{Huang2012b,Anderson2016}). The most ``distinct" signal found in our approach is a $\approx 43$ GeV line with a local significance of $\sim 4.1\sigma$ (see Fig.\ref{fig_sed_p8r2} and the left panel of Fig.\ref{slideWindowP8R2}). After the trial factor correction (see Fig.\ref{10sim}) the significance reduces to $\sim 3.0\sigma$. The analysis of the Earth limb data does not reveal a similar signal (see the right panel of Fig.\ref{slideWindowP8R2}). If the line signal can be confirmed by future data and the instrumental origin can be convincingly ruled out, it will have some interesting implications: (1) the boost factor due to the dark matter substructures of the Galaxy clusters should be in the order of ${BF} \geq  10^{3}$, as found in some simulations \cite{Gao2012}, otherwise it will be in contradiction with the constraints set by the current Galactic gamma-ray data \cite{Albert2015Line}; (2) the DM distribution in the inner Galaxy should be isothermal-like, otherwise the required ${BF}$ is too large (i.e., $\sim 10^{4}$) to be favored.

Since the global significance of the line signal is relatively low and the instrumental effects are to be better explored, we have estimated the upper limits on the DM annihilation as a function of $m_\chi$. Such constraints are much weaker than that set by the Galactic emission data (see Fig.\ref{fig_ul_anni}) , consistent with that found before (e.g. \cite{Huang2012b,Anderson2016}). Finally, we would like to point out that Dark Matter Particle Explorer \cite{Chang2014}, a Chinese space mission dedicated to measure high energy cosmic ray and gamma-rays with unprecedented energy resolution in a wide energy range, is expected to considerably increase the sensitivity of the gamma-ray line search.

\begin{acknowledgments}
We thank the anonymous referee for helpful comments and R. Z. Yang, J. N. Zhou and B. Zhou for their collaboration on the line search in galaxy clusters in 2013. We also appreciate  S. Li, N. H. Liao, Z. Q. Xia, Y. L. Xin, and Q. Yuan for cross-checking our SED presented in Fig.1. This work was supported in part by the National Basic Research Program of China (No. 2013CB837000), National Natural Science Foundation of China under grants No. 11525313 (i.e., the Funds for Distinguished Young Scholars) and No. 11103084, by Foundation for Distinguished Young Scholars of Jiangsu Province, China (No. BK2012047), and by the Strategic Priority Research Program (No. XDA04075500) of Chinese Academy of Sciences.
\end{acknowledgments}

$^\ast$Corresponding authors (xiangli@pmo.ac.cn, yzfan@pmo.ac.cn, huangxiaoyuan@gmail.com, chang@pmo.ac.cn).

\end{document}